# MAGNETIC PROPERTIES of GaMnAs SINGLE LAYERS and GaInMnAs SUPERLATTICES INVESTIGATED AT LOW TEMPERATURE AND HIGH MAGNETIC FIELD

C. Hernandez[1], F. Terki[1], S. Charar[1], J. Sadowski[2,3,4*], D. Maude[5], V. Stanciu[6], P. Svedlindh[6]

[1]Groupe d'Etude des Semiconducteurs CC074, Université Montpellier II, Place Eugène Bataillon, 34095 Montpellier Cedex 5, France
[2]Niels Bohr Institute, Copenhagen University, DK-2100 Copenhagen, Denmark
[3]Institute of Physics, Polish Academy of Sciences, al. Lotnikow 32/46, PL-02-668 Warszawa, Poland
[4]MAX-Lab, Lund University, PO. Box 118, 221 00 Lund, Sweden
[5]High Magnetic Field Laboratory, CNRS-MPI, 25 Avenue des Martyrs, 38042 Grenoble, France.
[6]Uppsala University, 751 20 Uppsala, Sweden.

Magnetotransport properties of GaMnAs single layers and InGaMnAs/InGaAs superlattice structures were investigated at temperatures from 4 K to 300 K and magnetic fields up to 23 T to study the influence of carriers confinement through different structures. Both single layers and superlattice structures show paramagnetic-to-ferromagnetic phase transition. In GaMnAs/InGaAs superlattice beside the Curie temperature ($T_c \sim 40$ K), a new phase transition is observed close to 13 K.

## 1. Introduction

Since the discovery of ferromagnetism in III-V diluted magnetic semiconductors, InMnAs in 1992 [1] and GaMnAs in 1996 [2], with the maximum critical temperatures ($T_c$) of 50 K and 110 K respectively, an increasing number of research groups started to study magnetic and transport properties of these materials for prototype spintronic devices. III-V Diluted magnetic semiconductors are promising materials for new semiconductor spin devices since they are compatible with III-V heterostructures and they offer the opportunity to study spin related phenomena produced by hole mediated ordering of the local Mn spins via the sp-d interaction of Ruderman-Kittel-Kasuya-Yosida (RKKY). Interesting phenomena have been found both in single layers and multilayer structures implementing these magnetic semiconductors. Moreover the temperature of paramagnetic to ferromagnetic phase transition in GaMnAs has been significantly raised, up to 150 - 160 K due to the specific post-growth annealing procedures [3, 4]. In this work, we focus on magnetotransport properties of GaMnAs thin layers (100 Å – 300Å) and short period superlattices with GaMnAs or InGaMnAs in magnetic layers and GaAs or InGaAs in non magnetic spacers.

---

[*] corresponding author: e-mail: sadow@maxlab.lu.se, fax: +46 46 222 47 10

**2. Results and discussion**

The samples were grown by Low Temperature Molecular Beam Epitaxy (LTMBE) technique. The In composition in InGaAs superlattice spacers and InGaMnAs magnetic layers was chosen to be equal to 50 % while the Mn concentration was equal to 5.5 % in GaMnAs single layers and 6 % in the magnetic layers of superlattice structures. The post growth annealing known to increase Tc and hole concentrations in III-Mn-V magnetic semiconductors was not used in this study. All the samples described here were not annealed intentionally after the MBE growth process, however due to the large number of repetitions (200) used for SL structures, the time of the MBE growth process was 8 – 12 h, depending on the structure. Growth rate of all the samples was the same and equal to 0.2 ML/s, in the case of SLs structures the growth interruptions at interfaces were 20 s – 40 s long. This, as we have shown elsewhere [5] improves both magnetic ($T_c$) and structural (intensities of X-ray diffraction Bragg satellite peaks) properties of SLs.

The thin (150 Å – 300 Å) GaMnAs single layers were covered by a 30 Å thick GaAs layer for protection against GaMnAs surface oxidation. Beside single GaMnAs thin films, two types of superlattices were studied, differing in band offsets between magnetic layers and nonmagnetic spacers: GaMnAs/GaAs and InGaMnAs/InGaAs SLs with very small band offsets between magnetic and non magnetic layers, and GaMnAs/InGaAs SLs with magnetic layers as potential barriers for carriers. Details of the epitaxial growth conditions of InGaMnAs/InGaAs SLs as well as basic structural properties were described elsewhere [6]. Magnetotransport measurements were carried out using Hall bars produced by UV lithography technique and chemical etching, with AuZn ohmic contacts. The temperature dependence of resistivity was studied from 4 K to 300 K for both single layers and SL structures. In short period SL structures magnetoresistance and Hall effect were investigated at magnetic fields up to 23 T and temperatures from 4 K to 270 K. Previous magnetization measurements performed by SQUID magnetometry show that all these samples present the paramagnetic to ferromagnetic phase transition with $T_c$ ranging from 10 K to 104 K depending on the sample [6, 7]. In the first set of samples, we have studied the resistivity temperature dependence of thin GaMnAs single layers with thicknesses of 100, 150, 200 and 300 Å. As shown in Fig. 1. the resistivity temperature dependence of those GaMnAs single layers indicates Curie temperatures of 104 K, 102 K, 93 K and 75 K respectively. The layers are on the metal side of the nonmetal-metal transition with a local maximum in resistivity at $T_c$ [8]. These results are in agreement with previous studies showing that $T_c$ depends on the thickness of the sample. It decreases with increasing the thickness [3, 7]. The second set of samples characterized by magnetotransport measurements are GaMnAs/GaAs, GaMnAs/InGaAs and InGaMnAs/InGaAs superlattice (SL) structures with very thin magnetic layers: 8 molecular layers (ML) (23 Å), and thin nonmagnetic spacers (4 – 8 ML). The magnetic layers thickness is below the thickness limit for ferromagnetism in a single GaMnAs layer (~ 5 nm) [7]. We detected the ferromagnetic transition in SLs with $T_c$ between 10 K and 65 K depending on the SL structure. Concerning the GaMnAs (8 ML) / GaAs (8 ML) x200 SL, one can observe that it has the same Hall resistance and resistivity behaviour as observed in a single GaMnAs layer. Anomalous Hall effect at low temperature reveal the conventional ferromagnetic ordering (see the inset in Fig.2).

In general the Hall resistance in ferromagnetic materials is described by the following equation,

$$R_{Hall} = \frac{R_0}{d}B + \frac{R_S}{d}M \,\, , (2.1)$$

where $R_0$ is the ordinary Hall coefficient, $R_s$ is the anomalous Hall coefficient, $M$ the magnetization of the sample and $d$ the thickness of conductive layers. From these measurements $T_c$ for GaMnAs(8 ML) /GaAs(8 ML) SL was estimated to be near 60 K, using Arrott plots. The resistivity measurements show the same $T_c$. As shown in Fig. 2 this sample shows a semiconducting behaviour with thermal activation below $T_c$. The second SL structure investigated is InGaMnAs (8 ML) / InGaAs (4 ML) x200 where previous SQUID magnetization measurements show a $T_c$ close to 10K [6]. This sample was to resistive for Hall measurements at low temperatures. Nevertheless it exhibits an interesting feature near 200 K. The Hall resistance shows a change of the Hall coefficient sign at low magnetic field (Fig. 3.). This behaviour was also observed in single InGaMnAs layers [9]. It has been suggested [9] that the different sign of the Hall coefficient is probably due to the spin-related scattering occurring above $T_c$ at external magnetic field. If this argument is true it is not clear why this effect was not observed in all the structures with high Mn content.

The last SL structure investigated is GaMnAs (8 ML) / InGaAs (4 ML) x200, with magnetic layers as potential barriers for carriers. Fig. 4 presents the temperature dependence of resistivity. It can be seen that in this structure, the paramagnetic to ferromagnetic transition which occurs at the Curie temperature (40 K) corresponds to an inflection point. This value is in good agreement with previous magnetization measurements [6]. From anomalous Hall effect (Fig.5.), we have deduced the same value of $T_c$ using an Arrott Plot [10]. These results indicate that the side-jump mechanism is responsible for the presence of the anomalous Hall effect. Assuming that magnetization of this sample at 20 K saturates at high magnetic fields, we have estimated $R_0$ and deduced the hole concentration ($p \sim 9.1 \times 10^{18}$ cm$^{-3}$). This is lower than the hole concentration in a single GaMnAs layer (~ $10^{20}$ cm$^{-3}$). The reason for that could be simply explained by the fact that the Mn doping is not present in the spacer layers. Moreover, at no external magnetic field, a sharp increase of the electrical resistivity with decreasing temperature occurs, and a bump is observed at 13 K. This feature indicates that a new phase with another type of ordering appears. It is the first evidence, by transport measurements of a new magnetic arrangement in SL structures. As predicted by Jungwirth et. al. [11] for GaMnAs/GaAs superlattices, the antiferromagnetic alignment of adjacent magnetic layers may appear at this temperature. A second explanation of this bump in the resistivity could be the Mn diffusion in non magnetic spacer with another ordering that the adjacent magnetic layers. Another reason of the 13 K magnetoresistivity maximum could be related to the confinement of carriers in non magnetic spacers. Such a transition has not been observed in single GaMnAs layers.

## Conclusions

We have studied magnetic properties of different kind of GaMnAs and InGaMnAs structures (single layers and superlattices) via resistivity and Hall measurements. These structures present a low temperature ferromagnetic phase transitions with $T_c$ within the range of 10K - 104K. Results for the GaMnAs/InGaAs SL, with magnetic layers as potential barriers for holes evidenced a new phase transition at the temperature lower than $T_c$. Further measurements such as neutron reflectivity, transmission electron microscopy (TEM) or scanning tunneling microscopy (STM) are needed to explain that. These complementary investigations will permit to detect the possible diffusion of Mn atoms in non magnetic spacers, and confirm/exclude the possibility of antiferromagnetic interlayer coupling as the origin of this new ordering.

**Acknowledgments**

Measurements at Grenoble High Magnetic Fields Laboratory have been supported by the European Community within the "Access to Research Infrastructure action of the Improving Human Potential Programme".

We are grateful to I. Salesse for performing AuZn contacts in Hallbar geometry using UV lithography technique and chemical etching at the ATEMI laboratory of Université Montpellier II.

**References**

[1] H. Ohno, H. Munekata, S. Von Molnar, L. L. Chang, Phys.Rev.Lett. **68**, 2664 (1992).

[2] H. Ohno, A.Shen, F. Matsukura, A. Oiwa, A. Endo, S. Katsumoto, Y. Iye, Appl.Phys.Lett. **69**, 363 (1996).

[3] K. C. Ku, S. J. Potashnik, R. F. Wang, S. H. Chun, P. Schiffer,
N. Samarth, M. J. Seong, A. Mascarenhas, E. Johnston-Halperin, R. C. Myers, A. C. Gossard, and D. D. Awschalom, Appl. Phys. Lett. **82**, 2302 (2003)

[4] K. W. Edmonds, K. Y. Wang, R. P. Campion, A. C. Neumann,
N. R. S. Farley, B. L. Gallagher, and C. T. Foxon, Appl. Phys. Lett. **81**, 4991 (2002)

[5] J. Sadowski, R. Mathieu, P. Svedlindh, M. Karlsteen, J. Kanski, Y. Fu,
J. Z. Domagała, W. Szuszkiewicz, B. Hennion, D. K. Maude, R. Airey, G. Hill,
Thin Solid Films **412**, 122 (2002)

[6] J. Sadowski R. Mathieu, P. Svedlindh, J. Kanski, M. Karlsteen, K Świątek,
J. Z. Domagała, Acta Phys. Pol. A **102**, 687 (2002).

[7] B.S. Sorensen, J. Sadowski, S. E. Andresen, and P. E. Lindelof,
Phys.Rev.B **66** 233313 (2002).

[8] A. Oiwa, S. Katsumoto, A. Endo, M. Hirasawa, H. Ohno, Y. Sugawara, A. Shen, F. Matsukura and Y. Iye, Solid State Commun. **103**, 209 (1997)

[9] T. Slupinski, H. Munekata, A. Oiwa, Appl.Phys.Lett. **80**, 1592 (2002).

[10] C. Hernandez, F. Terki, S. Charar, J. Sadowski, D. Maude, J. Magn. Mag. Mat. – in press (2003).

[11] T. Jungwirth, W. A. Atkinson, B. H. Lee, A. H. MacDonald, Phys.Rev.B **59**, 9818 (1999).

**Figure captions:**

**Fig. 1.** Temperature dependence of the reduced resistivity $(\rho-\rho_0)/\rho_0$ in $Ga_{0.945}Mn_{0.055}$ As single layers: 100Å - 150Å - 200Å - 300Å thick. $\rho_0$ is the resistivity at 4K.

**Fig. 2.** Temperature dependence of resistivity $\rho$ of the $Ga_{0.94}Mn_{0.06}As$(8 ML)/GaAs(8 ML) x 200 superlattice at zero magnetic field. At low magnetic field anomalous Hall effect is preponderant (insert).

**Fig. 3.** Hall resistance measurements of $(In,Ga)_{0.94}Mn_{0.06}As$(8 ML)/$In_{0.5}Ga_{0.5}As$(4 ML) x 200 SL. At high temperature ( > 200 K) the Hall coefficient sign changes at low magnetic field.

**Fig. 4.** Temperature dependence of resistivity ρ of the superlattice $Ga_{0.94}Mn_{0.06}As$(8 ML)/$In_{0.5}Ga_{0.5}As$(4 ML) x 200 SL at zero field. The inset shows the effect of magnetic field below $T_c$.

**Fig. 5.** Hall resistance measurements of $Ga_{0.94}Mn_{0.06}As$(8 ML)/$In_{0.5}Ga_{0.5}As$(4 ML) x 200 SL with temperature as a parameter. The inset shows the linear behaviour of Hall resistance at high magnetic field, from which the hole concentration was calculated.

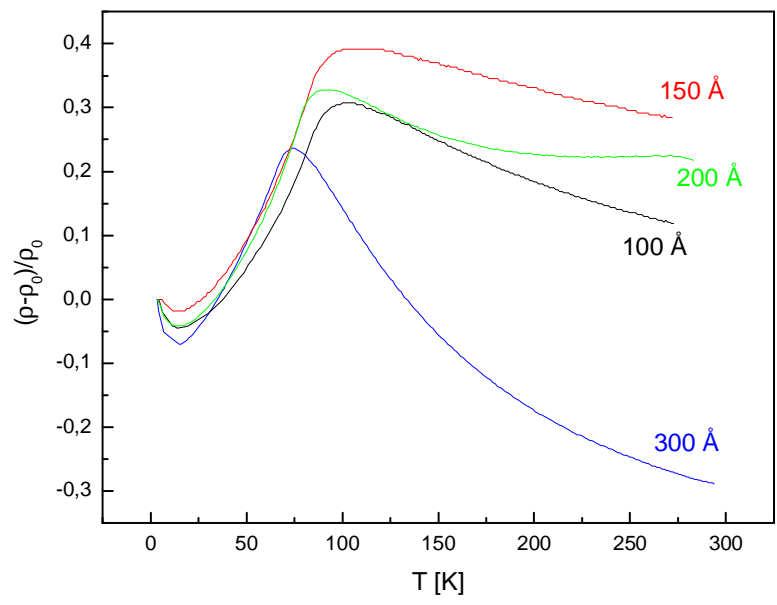

C. Hernandez et. al.   Fig. 1.

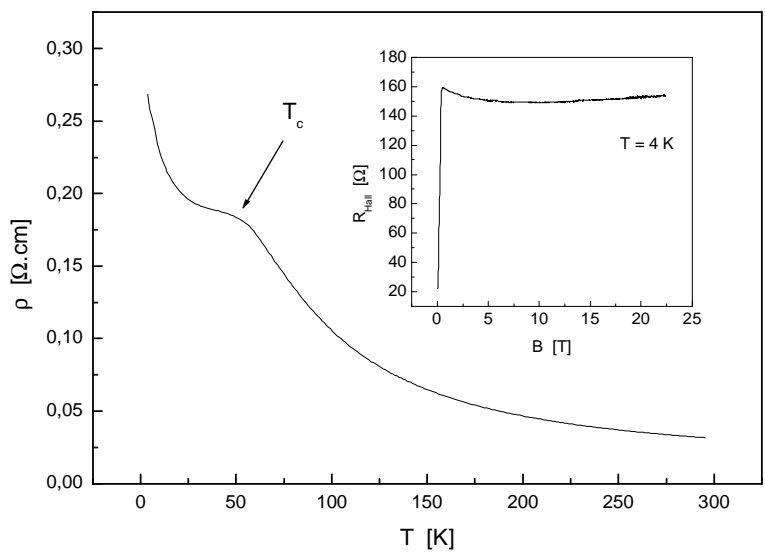

C. Hernandez et. al.   Fig. 2.

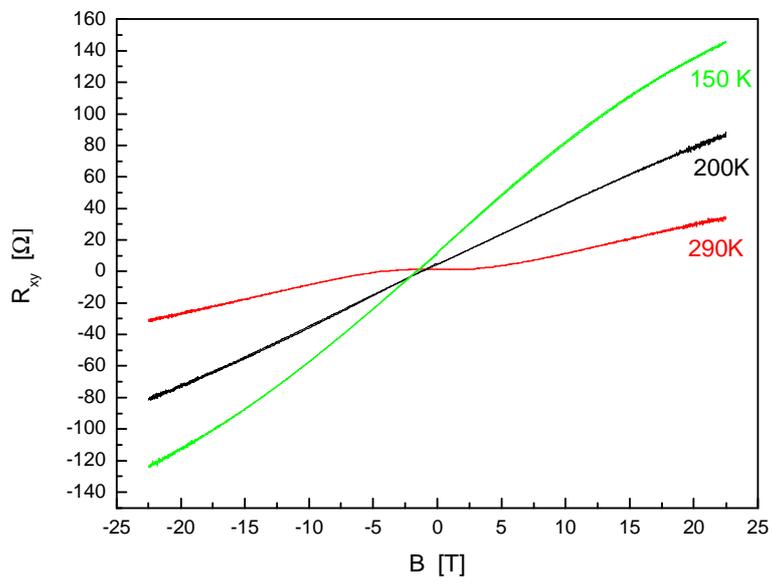

C. Hernandez et. al.   Fig. 3.

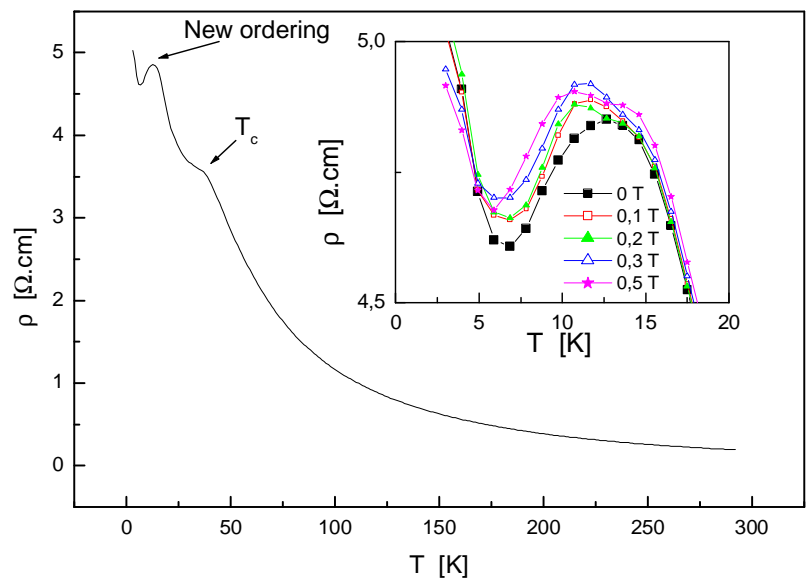

C. Hernandez et. al. Fig. 4.

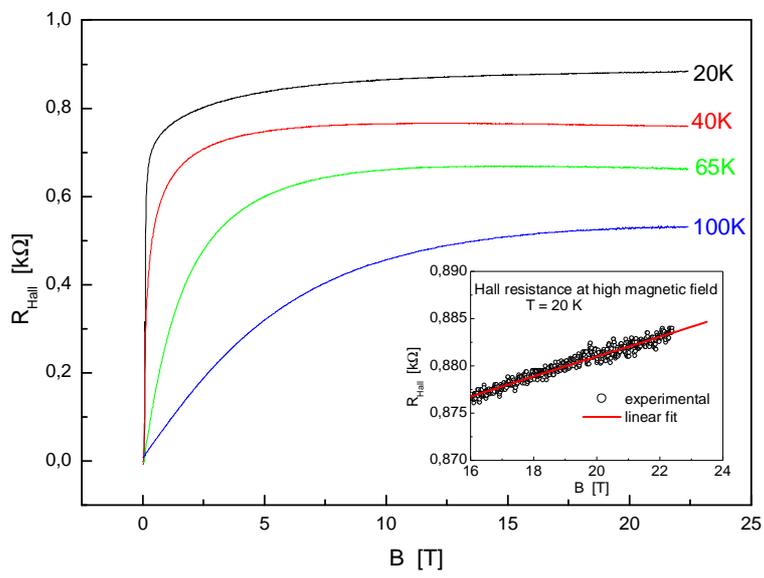

C. Hernandez et. al.   Fig. 5.